\documentclass[aps,longbibliography,twocolumn,superscriptaddress]{revtex4-2}

\usepackage[utf8]{inputenc}
\usepackage[T1]{fontenc}
\usepackage{amsmath}
\usepackage{amssymb}
\usepackage{amsfonts}
\usepackage{bm}
\usepackage{color}
\usepackage{xcolor}
\usepackage{datetime}
\usepackage{graphicx}
\usepackage[colorlinks,citecolor=blue,urlcolor=blue,linkcolor=blue]{hyperref}
\usepackage{braket}
\usepackage{mathtools}
\usepackage[colorinlistoftodos]{todonotes}

\newcounter{chcount}
\definecolor{ggg}{rgb}{0.0, 0.5, 0.0}

\newcommand{\Tr}{\mathrm{tr}}
\newcommand{\inext}{{}_\mathrm{ext}}
\newcommand{\inint}{{}_\mathrm{int}}

\begin{document}

\title{Many-body coherence and entanglement probed by randomized correlation measurements}

\author{Eric Brunner}
\email{eric.brunner@physik.uni-freiburg.de}
\affiliation{Physikalisches Institut, Albert-Ludwigs-Universit{\"a}t Freiburg, Hermann-Herder-Str. 3, 79104 Freiburg, Germany}
\affiliation{EUCOR Centre for Quantum Science and Quantum Computing, Albert-Ludwigs-Universit{\"a}t Freiburg, Hermann-Herder-Str. 3, 79104 Freiburg, Germany}

\author{Andreas Buchleitner}
\affiliation{Physikalisches Institut, Albert-Ludwigs-Universit{\"a}t Freiburg, Hermann-Herder-Str. 3, 79104 Freiburg, Germany}
\affiliation{EUCOR Centre for Quantum Science and Quantum Computing, Albert-Ludwigs-Universit{\"a}t Freiburg, Hermann-Herder-Str. 3, 79104 Freiburg, Germany}

\author{Gabriel Dufour}
\affiliation{Physikalisches Institut, Albert-Ludwigs-Universit{\"a}t Freiburg, Hermann-Herder-Str. 3, 79104 Freiburg, Germany}
\affiliation{EUCOR Centre for Quantum Science and Quantum Computing, Albert-Ludwigs-Universit{\"a}t Freiburg, Hermann-Herder-Str. 3, 79104 Freiburg, Germany}



\begin{abstract}
We show how coherences between identical constituents of a many-body quantum state can be interrogated by suitable correlation 
functions, and identify sufficient conditions under which low-order correlators fully characterize many-body coherences, as controlled by the constituents' mutual distinguishability. Comparison of correlators of different order detects many-body entanglement.
\end{abstract}

\maketitle

Coherence properties of many-body quantum states are essential, e.g. for purposes of quantum information \cite{nielsen-chuang} or control \cite{scholes2017}, as well as in elementary scattering processes \cite{Tannoudji-QM-1973} of photons \cite{hong_measurement_1987}, electrons \cite{mott1930,saraga2004} or protons \cite{mott1930,segre1977}. For larger particle numbers, though, they are hard to characterize \cite{haeffner2005}, due to the unfavourable scaling properties of state space with the number of constituents. On the other hand, many-body interference (MBI) opens up an entirely new realm of rich, multi-facetted interference phenomena \cite{Tannoudji-QM-1973,hong_measurement_1987,tichy_zero-transmission_2010} beyond conventional single particle interference probed, e.g., in the center of mass degree of freedom (dof) of composite quantum objects \cite{arndt2005}. MBI also establishes another perspective upon the quantum-classical transition -- here controlled by the constituent identical particles' mutual level of distinguishability \cite{hong_measurement_1987,mayer_counting_2011,ra_nonmonotonic_2013,dittel_wave-particle_2019} rather than, e.g., by their accumulated mass \cite{arndt2005,schlosshauer2019} or total number \cite{kolovsky2003,hiller2006,rammensee2018,preiss2020}. Experimental tools have now reached a level of sophistication which allows to prepare and interrogate many-body states with unprecedented control of the number of constituents, as well as of their external (acted upon, e.g., by optical potential landscapes) and internal (defined, e.g., by a single particle's electronic states) dof \cite{tillmann_generalized_2015,menssen_distinguishability_2017,agne_observation_2017,jones_multiparticle_2020,pleinert_2021,zache_extracting_2020,shadbolt_generating_2012,russell_direct_2017,mennea_modular_2018,sansoni_two-particle_2012,matthews_observing_2013,preiss2020}. In turn, experiments also clearly witness the enhanced fragility of many-body coherences with increasing particle number \cite{google2019}, while a full-fledged theory of many-body (de-)coherence is still in the making.

In this general context, it is necessary to understand which observables are well-suited to distil distinctive target properties of a given resource state while warranting
benign experimental overhead with respect to scaling with the particle number. Since, by the very nature of complex quantum systems, it is also clear that such observables can never exhaustively characterise a given state's properties (think, e.g., of universal vs. system-specific features characterised by random matrix vs. semiclassical theories of chaotic quantum systems \cite{leshouches1989,welge1988,delande1986,rammensee2018,pausch2021}), we further need a precise understanding of those potentially relevant system properties which a given observable is blind to.

A particularly transparent setting to proceed in this direction is offered by systems of non-interacting identical particles (such as photons, or suitably tuned \cite{gustavsson2008} 
cold atoms), equipped with internal dof (such as polarisation, arrival time, or an electronic dof) which allow to tune their level of mutual distinguishability \cite{hong_measurement_1987,mayer_counting_2011,walschaers_many-particle_2016}. When submitted to a unitary evolution in their (external) motional dof, as mediated, e.g., by a multi-mode scatterer, the many-body output will typically exhibit strong MBI contributions, arising from the  many-body coherence of the initial state. These interferences, however, will fade away as the particles acquire a finite level of distinguishability, via preparation in distinct 
states of their internal dof.

While specific output event probabilities \cite{tichy_zero-transmission_2010,dittel_totally_2018} or statistical features of low-order correlations \cite{walschaers_statistical_2016,walschaers_many-particle_2016,rigovacca_nonclassicality_2016,brod_witnessing_2019,giordani_experimental_2020,giordani_witnesses_2021,van_der_meer_experimental_2021,galvao_quantum_2020}
often are sensitive probes of MBI, without the necessity to record the full output statistics, it remained hitherto unclear which specific properties of the state under scrutiny are probed by these quantifiers, and, in turn, which many-body coherence properties may go undetected. 

We close this gap by systematically identifying orders $k=2, \dots N$ of many-body coherence in states of $N$ partially distinguishable (PD) bosons or fermions that control the MBI contributions to $k$-particle ($k$P) measurements.
We propose a quantifier of $k$P coherence and describe a protocol for its estimation based on an average over $k$-point correlation functions,
allowing for an order by order characterization of a state's many-body coherence.
By relating $k$P coherence to many-body distinguishability and entanglement, we identify conditions under which low-order correlators convey all essential information.

\paragraph*{Partially distinguishable particles ---}
Many-body states of \textit{partially distinguishable} particles are represented in the bosonic or fermionic Fock space $\mathcal{F}[\mathcal{H}]$ erected upon a single-particle (1P) Hilbert space describing both external and internal dof: $\mathcal{H} = \mathcal{H}\inext \otimes \mathcal{H}\inint$ \cite{adamson_multiparticle_2007,adamson_detecting_2008,stanisic_discriminating_2018,dittel_wave-particle_2019}. In contrast to the former, we assume that the latter are neither affected by the dynamics, nor interrogated by measurement, but only allow to (partially) distinguish the particles. The distinction between external and internal dof, and the associated notion of PD and entanglement are not absolute but determined by the experiment.
A basis of many-body states is provided by Fock states created from the vacuum $\ket{0}$ by multiple application of (bosonic or fermionic) creation operators $\hat{a}^\dagger_{p\alpha}$, where Latin (Greek) indices $p \in \mathcal{B}\inext$ $(\alpha\in\mathcal{B}\inint)$ label orthogonal external (internal) basis modes. For simplicity, we take $\mathcal{H}\inext$ and $\mathcal{H}\inint$ to be finite-dimensional and set $d=\dim\mathcal{H}\inext$. For a chosen basis $\mathcal{B}\inext$, Fock space $\mathcal{F}[\mathcal{H}]$ can be decomposed into the tensor product of Fock spaces built upon the internal dof, each associated with one of the orthogonal external modes $p\in\mathcal{B}\inext$: $\mathcal{F}[\mathcal{H}] \simeq \bigotimes_{p\in\mathcal{B}\inext} \mathcal{F}_p[\mathcal{H}\inint]$. A state is \textit{separable} in those external modes, in short \textit{externally separable}, if it is separable according to this partition. Otherwise, it is called \textit{externally entangled} \cite{benatti_entanglement_2020}.

The \textit{external number operator} $\hat{N}_{p} = \sum_{\alpha\in\mathcal{B}\inint} \hat{a}^\dagger_{p\alpha}\hat{a}_{p\alpha}$ counts the number of particles in mode $p \in\mathcal{B}\inext$, irrespective of their internal states.
In the following, we assume that the $N$-particle ($N$P) state $\rho$ whose coherence we want to characterize, e.g. the state generated by a many-particle source in a MBI experiment, can be prepared such that $N_p\in\lbrace 0,1\rbrace$, while imposing 
\textit{no condition on the structure of the many-particle state in its internal dof}, which we would like to assess.
Such a setting allows for a particularly transparent analysis of the interdependence of PD, coherence and entanglement \cite{brunner_many-body_2019} and, further, has come into reach of experiment, on diverse platforms. Pure, externally separable states with $N_p \in \lbrace 0,1 \rbrace$ are precisely those states where each particle occupies a distinct external mode $p_i$
and carries an individual, arbitrary, pure internal state 
$\ket{\phi_i} =\sum_\alpha \phi_i^\alpha \ket{\alpha} \in \mathcal{H}\inint,\, i = 1,\dots , N$ \cite{brunner_many-body_2019}.
Note, however, that many-particle sources, based e.g. on spontaneous parametric down conversion or quantum dots, can also be used to generate externally entangled states.

\paragraph*{Mode correlations and reduced states ---}

For an $N$P input state $\rho$, we consider non-interacting dynamics in the external modes (e.g., a linear interferometer), such that the evolution operator $\mathcal{U}$ acts as $\mathcal{U}^\dagger \hat{a}^\dagger_{p\alpha} \mathcal{U} = \sum_{m\in\mathcal{B}\inext} U_{pm} \hat{a}^\dagger_{m\alpha}$, with $U$ a unitary transformation on $\mathcal{H}\inext$. MBI in the external modes is assessed through measurements of many-body observables that are blind to the internal dof \cite{dufour_many-body_2020}. Typical examples are density correlation measurements between $k\leq N$ modes
\begin{equation}\label{eq:kPC}
\begin{split}
&\text{tr}\big[\rho\, \mathcal{U}^\dagger \hat{N}_{p_1} ... \hat{N}_{p_k} \mathcal{U}\big]
\\
&\quad= \sum_{ \bm{m},\bm{n}\in\mathcal{B}\inext^k  } 
\prod_{i=1}^kU_{p_im_i} U^*_{p_in_i}
\sum_{\bm{\alpha}\in\mathcal{B}\inint^k}
\text{tr}\left[ \rho\, \hat{a}_{\bm{m}\bm{\alpha}}^\dagger \hat{a}_{\bm{n}\bm{\alpha}} \right] 
\end{split}
\end{equation}
with orthogonal $p_i\in\mathcal{B}\inext$, $\hat{a}^{\dagger}_{\bm{m}\bm{\alpha}} = \hat{a}^{\dagger}_{m_1\alpha_1}\dots \hat{a}^{\dagger}_{m_k\alpha_k}$, and $\hat{a}_{\bm{m}\bm{\alpha}} = (\hat{a}^\dagger_{\bm{m}\bm{\alpha}})^\dagger$, for multi-indices $\bm{m}=(m_1, \dots m_k) \in \mathcal{B}\inext^k ,\, \bm{\alpha}=(\alpha_1, \dots \alpha_k) \in \mathcal{B}\inint^k$. The external $k$th-order correlation functions $\sum_{\bm{\alpha}\in \mathcal{B}\inint^k} \text{tr}[\rho\, \hat{a}^\dagger_{\bm{m}\bm{\alpha}} \hat{a}_{\bm{n}\bm{\alpha}} ]$ appear already in \cite{glauber_quantum_1963} in the study of coherence in many-body systems. They can be identified \cite{brunner_many-body_2019} with the matrix elements $\braket{\bm{n}| \rho\inext^{(k)} |\bm{m}} = \sum_{\bm{\alpha}\in \mathcal{B}\inint^k} \text{tr}[\rho\, \hat{a}^\dagger_{\bm{m}\bm{\alpha}} \hat{a}_{\bm{n}\bm{\alpha}} ] (N-k)!/N!$ of the \textit{external $k$P reduced density operator} $\rho\inext^{(k)}$ in the un-symmetrized (first quantization) product basis $\ket{\bm{m}} = \ket{m_1}\otimes \dots \otimes \ket{m_k}$ of $\mathcal{H}\inext^{\otimes k}$. Here, $\rho\inext^{(k)}$ is obtained from $\rho$ by embedding the $N$P Fock sector into $\mathcal{H}^{\otimes N} \simeq \mathcal{H}\inext^{\otimes N} \otimes \mathcal{H}\inint^{\otimes N}$ and performing the partial trace operations $\mathcal{H}\inext^{\otimes N} \otimes \mathcal{H}\inint^{\otimes N} \stackrel{\text{tr}\inint}{\longrightarrow} \mathcal{H}\inext^{\otimes N}  \stackrel{\text{tr}_{(N-k)}}{\longrightarrow} \mathcal{H}\inext^{\otimes k}$. Note that these traces commute \cite{brunner_many-body_2019,NoteX}. The $k$-point correlator \eqref{eq:kPC} therefore only accesses the $k$P marginal $\rho\inext^{(k)}$ of the initial many-body state $\rho$, and discards information collectively carried by larger numbers of particles.

The off-diagonal elements $\braket{\bm{n}| \rho\inext^{(k)} |\bm{m}},\, \bm{m}\neq \bm{n}$, are the \textit{$k$P coherences}. In Eq. \eqref{eq:kPC}, these come with weights defined by the specific unitary $U$. We show below that randomly chosen unitaries $U$, in combination with a suitable truncation scheme of the observable, realize (on average) an unbiased sampling of the $\braket{\bm{n}| \rho\inext^{(k)} |\bm{m}}$. This gives direct experimental access to the coherence of the initial state, and therefore of its capacity to display MBI, as quantified by the cumulative measures
\begin{equation}\label{eq:kpart_meancoh}
	\mathcal{W}^{(k)} = \sum_{\bm{m},\bm{n} \in\mathcal{B}\inext^k} \braket{\bm{n}| \rho\inext^{(k)} |\bm{m}} \,,
\end{equation}
which we baptize the \textit{$k$P mean coherence}. Hermiticity and positivity of $\rho\inext^{(k)}$ ensure that $\mathcal{W}^{(k)}$ is real and positive. 

\paragraph*{External separability and coherence ---}

For pure externally separable states, the non-zero matrix elements of $\rho\inext^{(k)}$ stem from multi-indices $\bm{m},\bm{n}$ that are connected via a \textit{unique} permutation $\pi\in\text{S}_k$ in the symmetric group of $k$ elements: $(m_1,\dots,m_k) = (n_{\pi^{-1}(1)},\dots, n_{\pi^{-1}(k)})$. They are given by products of overlaps of internal 1P states $\braket{\bm{n}| \rho\inext^{(k)} |\bm{m}} \propto \text{sgn}(\pi) \prod_{i=1}^k \braket{\phi_{m_i} | \phi_{n_i}}$, with $\text{sgn}(\pi)$ the signature of $\pi$ for fermions, and one for bosons.

In the classical limit 
\cite{NoteY} of perfectly distinguishable particles in mutually orthogonal internal states, the reduced density matrices $\rho\inext^{(k)}$ are diagonal, with $\mathcal{W}^{(k)}=\Tr\rho\inext^{(k)}=1$ at any order $k$.
Hence, in Eq. \eqref{eq:kPC} only diagonal terms $(\bm{m}=\bm{n})$ contribute and the measurement does not show any many-body interference signal. In turn, any deviation of $\mathcal{W}^{(k)}$ from one signals the existence of coherences in $\rho\inext^{(k)}$, giving rise to \textit{$k$P interference} contributions in Eq. \eqref{eq:kPC}. This extends the conventional interpretation of interference to the many-body setting. Indistinguishable bosons exhibit the maximum value of $\mathcal{W}^{(k)} = k!$, because all non-vanishing matrix elements of $\rho\inext^{(k)}$ are positive and equal. For indistinguishable fermions, each matrix element contributing to $\mathcal{W}^{(k)}$ is canceled by another one (due to the factor $\text{sgn}(\pi)$), resulting in $\mathcal{W}^{(k)}=0$.

Since 2P coherences are given by $\braket{m,n|\rho\inext^{(k)} | n,m} \propto \pm |\braket{\phi_{m}| \phi_{n}}|^2$ ($+$ for bosons and $-$ for fermions), $\mathcal{W}^{(2)}$ has a direct physical interpretation in terms of the particles' distinguishability, controlled by the overlaps of their internal states.
\begin{figure}
	\centering
	\includegraphics[width=\linewidth]{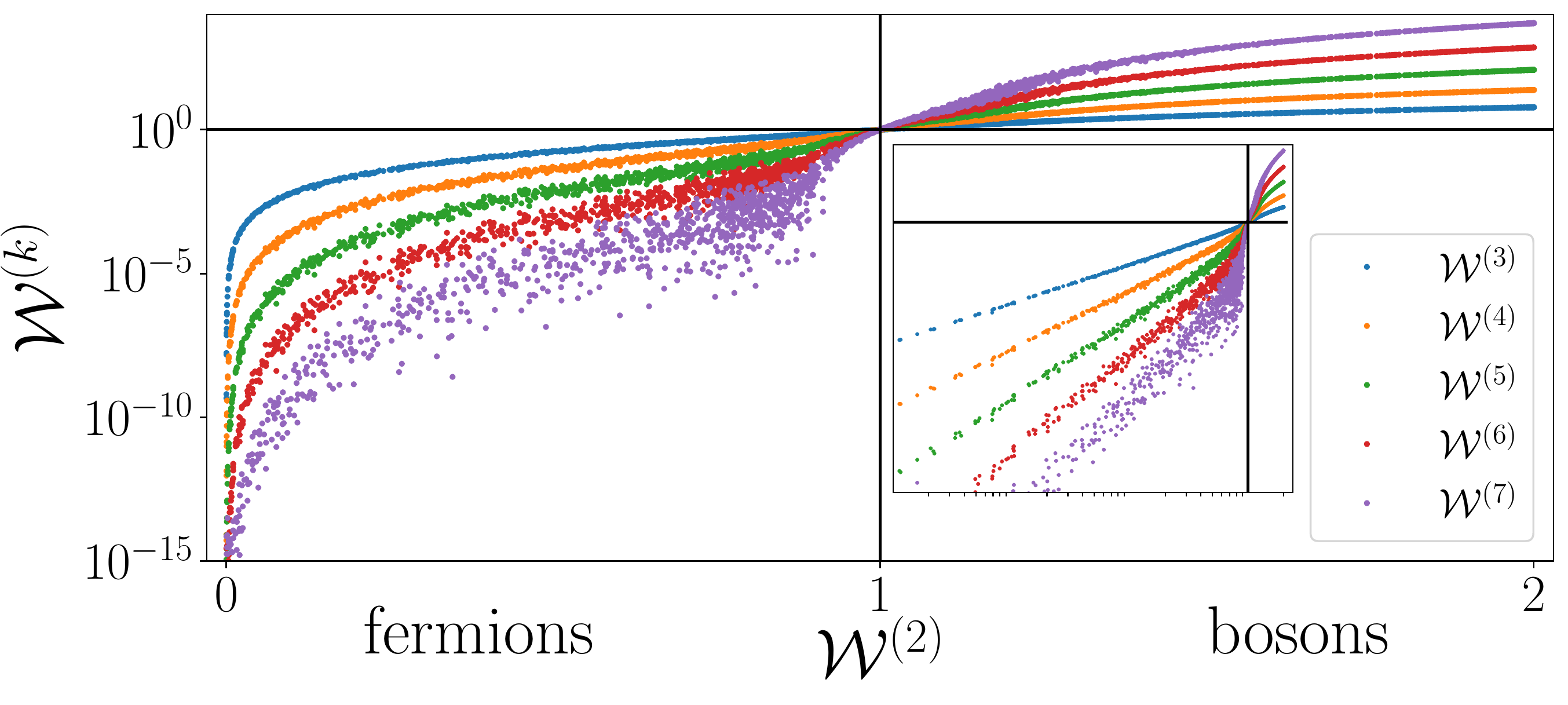}
	\caption{
		Correlation between $\mathcal{W}^{(k)}$ (log scale) and $\mathcal{W}^{(2)}$ from Eq.~(\ref{eq:kpart_meancoh}), $k=3,\dots7$,	for each of 1000 fermionic and bosonic externally separable 7P states on seven external modes, with each particle in a random internal pure state $\ket{\phi_j}$ (the detailed sampling procedure is supplemented \cite{SM}).	On average, $\mathcal{W}^{(k)}$ depends strictly monotonically on $\mathcal{W}^{(2)}$ over the entire range from indistinguishable fermions, $\mathcal{W}^{(2)}=0$,	to bosons, $\mathcal{W}^{(2)}=2$. All measures unambiguously discriminate	fermions, $\mathcal{W}^{(k)}<1$, distinguishable particles, $\mathcal{W}^{(k)}=1$, and bosons, $\mathcal{W}^{(k)}>1$. The inset shows the data in a double logarithmic plot. In the limit of indistinguishable fermions ($\mathcal{W}^{(2)} \rightarrow 0$) we empirically identify a power-law relation with exponent $k-1$.}
	\label{fig:meancoh_bosfer}
\end{figure}
Numerical analysis shows that, for externally separable states, higher-order mean coherences $\mathcal{W}^{(k)}$ are, in good approximation, given by monotonically increasing functions of $\mathcal{W}^{(2)}$. In Fig. \ref{fig:meancoh_bosfer}, we present scatter plots of $\mathcal{W}^{(k)},\, k>2,$ against $\mathcal{W}^{(2)}$, for states of seven particles in seven external modes, each with a randomly sampled (pure) internal state $\ket{\phi_i}\in\mathcal{H}\inint, \ i=1,\dots 7$, such as to cover the range of $\mathcal{W}^{(2)} \in [0,2]$ as uniformly as possible \cite{SM}. For  $k = 3,\dots ,7$, the simulation indicates that all $\mathcal{W}^{(k)}, k=2,\dots,7$, map out essentially the same transition from indistinguishable fermions ($\mathcal{W}^{(k)}=0$) to indistinguishable bosons ($\mathcal{W}^{(k)}=k!$), via distinguishable particles ($\mathcal{W}^{(k)}=1$).
On a log-log scale, we observe a power-law behavior in the limit of indistinguishable fermions with an exponent $k-1$ (see inset of Fig. \ref{fig:meancoh_bosfer}), which remains to be elucidated.

Note that, since the $\mathcal{W}^{(k)}$ are linear in the density matrix, the same, strictly monotonic relationship between the $\mathcal{W}^{(k)}$ holds for \textit{mixed} externally separable states. Moreover, the convex structure of mixed states will typically reduce the scatter, since deviations (if uncorrelated) will cancel out on average. {\it For externally separable states, $\mathcal{W}^{(k)}$ thus faithfully reflects the involved particles' PD, over the  entire range from indistinguishable fermions to indistinguishable bosons, via the intermediate case of distinguishable particles.} This generalizes the intimate connection between coherence and indistinguishability of single-particle paths \cite{mandel_coherence_1991}
to many-body systems.

Various quantities already considered in the literature fall within the framework of the $k$P coherence measures defined in Eq. \eqref{eq:kpart_meancoh}, albeit only for the extreme cases $k=2$ and $k=N$.
The \textit{degree of indistinguishability} $\mathcal{I}$, introduced in \cite{brunner_signatures_2018,dufour_many-body_2020} to quantify PD of bosonic Fock states, is proportional to $\mathcal{W}^{(2)}-1$, see \cite{SM}. 
The witness of \textit{genuine $N$-photon indistinguishability}, as considered in \cite{brod_witnessing_2019,giordani_experimental_2020,giordani_witnesses_2021}, is based on pairwise overlaps of the particles' internal states and can be rephrased \cite{SM} in terms of $\mathcal{W}^{(2)}$: Violation of $\mathcal{W}^{(2)} \leq 2-2/N$ implies \textit{genuine $N$-photon indistinguishability} in the above sense.
The $J$ matrix of \cite{shchesnovich_partial_2015,shchesnovich_tight_2015} is, in essence, our density matrix $\rho\inext$, but accounts in addition for possibly imperfect particle detection. 
In \cite{dittel_wave-particle_2019}, sums over all matrix-elements of the full external $N$P state $\rho\inext$, as in $\mathcal{W}^{(N)}$, are considered as a measure of PD, but their absolute value (squared) is taken,
which has the effect of erasing the difference between bosonic and fermionic statistics. 
Finally, $\mathcal{W}^{(N)}\prod_{m\in\mathcal{B}\inext} N_m!/N!$ measures the projection of $\rho\inext$ on the symmetric subspace of $\mathcal{H}\inext^{\otimes N}$, a quantity considered in \cite{dittel_wave-particle_2019,minke_characterizing_2021}, which also coincides with the \emph{degree of interference} of \cite{tichy_sampling_2015}. 
\textit{Our proposed definition of $\mathcal{W}^{(k)}$ links all these quantities to the coherence of the reduced states $\rho\inext^{(k)}$ \cite{SM} and allows for a direct interpretation in terms of various orders of $k$P interference processes.}

\paragraph*{Connected correlators and random matrix average---}
We now turn to the estimation of the $\mathcal{W}^{(k)}$ for a \textit{general}, i.e. possibly externally entangled, input state $\rho$. Since low-order interference terms in \eqref{eq:kPC} typically dominate the expectation value, we enhance higher-order contributions by employing the \textit{connected}, or \textit{truncated}, $k$-point correlators, recursively defined as
\begin{equation}\label{eq:connected}
\mathcal{C}^{(k)}_{\bm{p}}= \text{tr} \big[ \rho\, \mathcal{U}^\dagger \hat{N}_{p_1} \dots \hat{N}_{p_k} \mathcal{U} \big] - \sum_{P \vdash \bm{p}} \prod_{\bm{q}\in P} \mathcal{C}^{(|\bm{q}|)}_{\bm{q}} \,,
\end{equation}
where the sum runs over all non-trivial partitions $P\vdash \bm{p}$ of modes $\bm{p}=\lbrace p_1,\dots,p_k\rbrace$ into disjoint subsets $\bm{q}$ of length $|\bm{q}|$, each being associated with a possible factorization of the correlator. For example, for $k=2$, $\mathcal{C}^{(2)}_{p_1p_2} = \braket{\hat{N}_{p_1} \hat{N}_{p_2}}-\braket{\hat{N}_{p_1}}\braket{\hat{N}_{p_2}}$ is the covariance. Connected correlators are commonly used in various fields of physics (notably also in the theoretical analysis of many-body quantum systems \cite{zache_extracting_2020}) and mathematics, where they are also known as joint cumulants.

By choosing $U$ at random from the Haar measure \cite{haar_massbegriff_1933} on the unitary group $U(d)$, we perform a correlation measurement in randomly chosen external modes. Integration of \eqref{eq:connected} over the unitary group returns the average connected correlator, with the help of (for orthogonal $p_i$) \cite{collins_integration_2006}
\begin{equation}\label{eq:haarintegration}
\overline{\prod_{i=1}^k U_{p_im_i} U^*_{p_in_i}} = \sum_{\pi\in \text{S}_k} \text{Wg}_d(\pi) \, \prod_{i=1}^k \delta_{m_{\pi(i)}, n_i}\,  .
\end{equation}
The overline indicates the Haar integration and $\mathrm{Wg}_d$ is the Weingarten function \cite{weingarten_asymptotic_1978,collins_integration_2006}, which only depends on the cyclic structure of the permutation $\pi\in\text{S}_k$ and on the external dimension $d$. For $k=2,3$ and unit filling factor, i.e. $N=d$, the truncation \eqref{eq:connected} of the correlators and the Haar average \eqref{eq:haarintegration} cooperate in exactly the right way to ensure that all matrix elements of $\rho\inext^{(k)}$ are uniformly weighted
\cite{brunner_many-body_2019}:
\begin{equation}\label{eq:2pointcor}
\overline{\mathcal{C}^{(2)}_{p_1 p_2}} = -\frac{\mathcal{W}^{(2)}}{N+1}
~,\quad \overline{\mathcal{C}^{(3)}_{p_1 p_2 p_3}} = \frac{2\mathcal{W}^{(3)}}{(N+1)(N+2)}~.
\end{equation}
Note that this result holds also for externally entangled states. However, the interpretation of $\mathcal{W}^{(k)}$ as a PD measure is only valid in the case of externally separable input states, as discussed above. Relaxing the assumption $N=d$ leads to similar expressions, where the various matrix elements of $\rho\inext^{(k)}$ acquire different weights depending on $d$ and $N$, as we will show in detail elsewhere.

\begin{figure}
	\includegraphics[width=\linewidth]{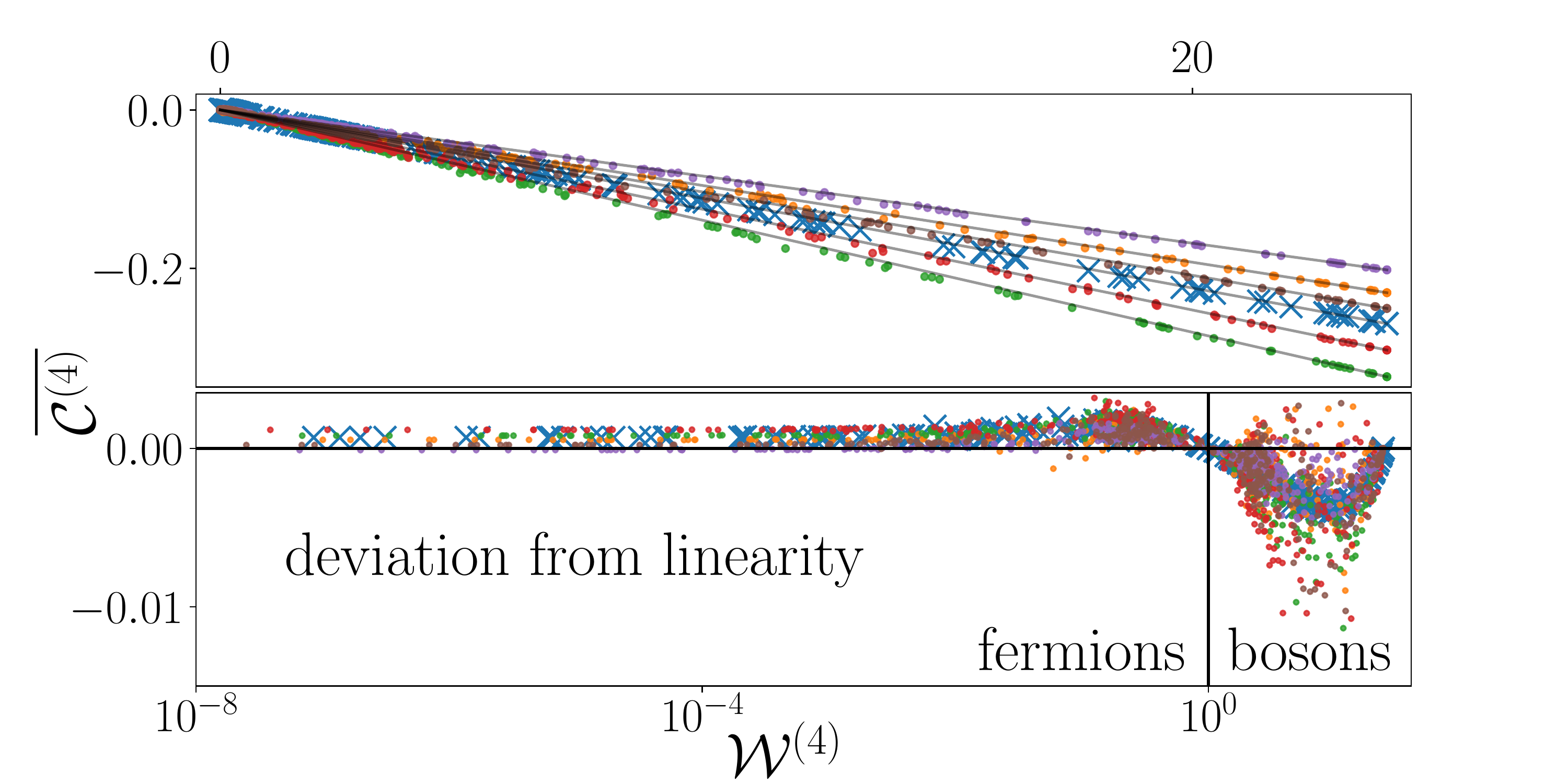}
	\includegraphics[width=\linewidth]{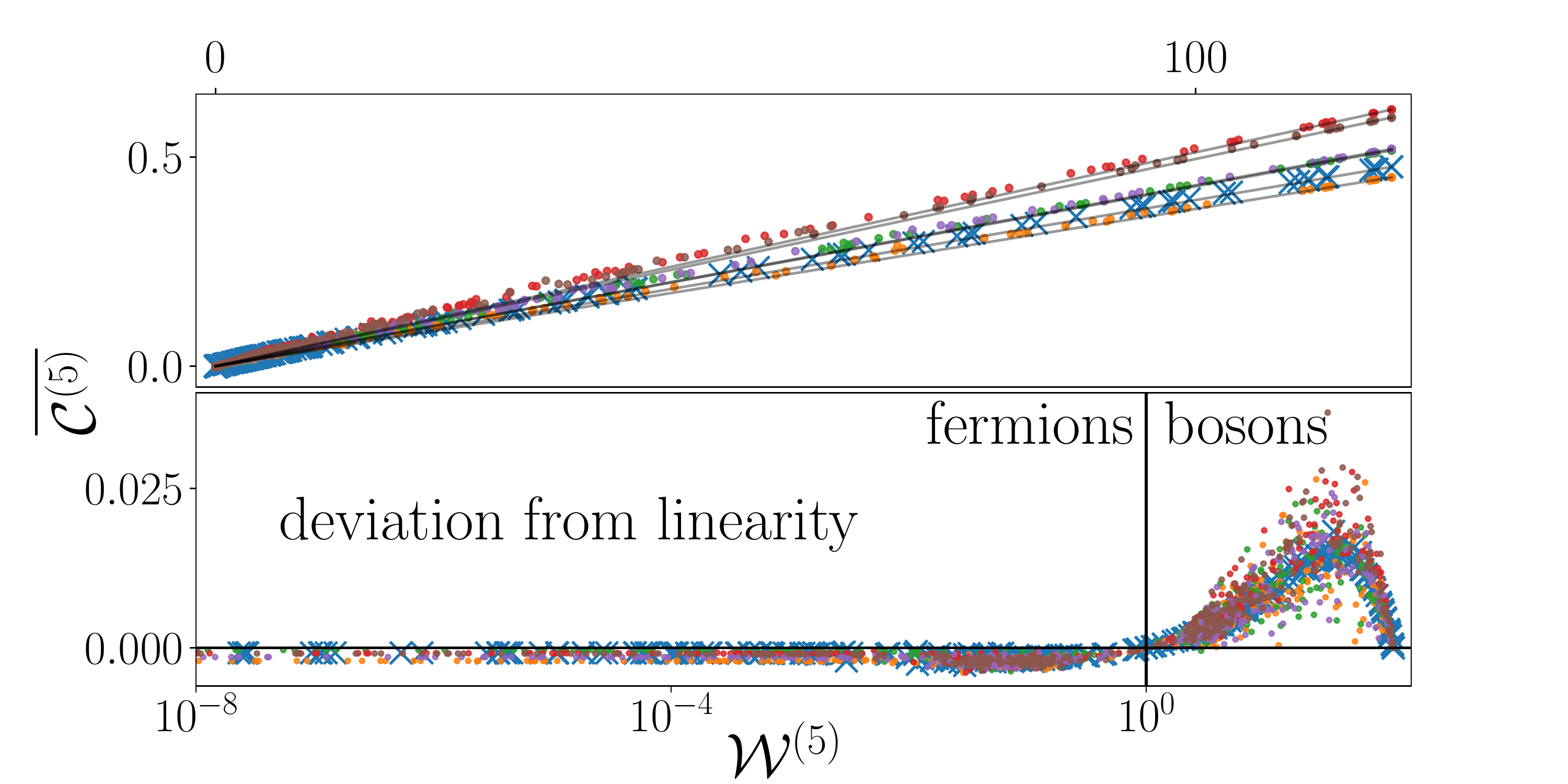}
	\caption{Connected correlators $\mathcal{C}^{(k)}_{p_1\dots p_k}$, eq.~\eqref{eq:connected}, 
	vs. $\mathcal{W}^{(k)}$, eq.~\eqref{eq:kpart_meancoh}, with $k=4$ (top) and $k=5$ (bottom), for six particles in six modes, averaged, for a given instance of a random Haar unitary, over all choices of $k$ out of $N$ output 	modes (coloured dots -- each colour represents another random unitary), compared to the random matrix prediction (blue crosses) obtained by integration over the Haar measure. The transition between indistinguishable fermions and bosons was covered by sampling random internal states $\ket{\phi_j}$ for each particle (as for Fig. \ref{fig:meancoh_bosfer} \cite{SM}). Bottom panels in each plot show the deviation of the data from a linear dependence of the averaged $\mathcal{C}^{(k)}_{p_1\dots p_k}$ on $\mathcal{W}^{(k)}$, from indistinguishable fermions ($\mathcal{W}^{(k)} = 0)$ to bosons ($\mathcal{W}^{(k)} = k!$). The logarithmic $x$-scale resolves the fermionic range ($0<\mathcal{W}^{(k)}<1$) in more detail.
	}
	\label{fig:corvsmeancoh}
\end{figure}

The linear relations \eqref{eq:2pointcor} do not exactly hold at higher correlation orders. However, as we show by numerical simulations, uniform sampling of the matrix elements of $\rho\inext^{(k)}$ in the input basis---through the introduced scheme of truncated randomized correlations---is observed, to very good approximation, also for $k>3$. In Fig. \ref{fig:corvsmeancoh}, we show that for 6P input states (sampled according to the same procedure as for Fig. \ref{fig:meancoh_bosfer} \cite{SM}), a tight relation persists between $\overline{\mathcal{C}^{(k)}_{p_1\dots p_k}}$ and $\mathcal{W}^{(k)}$ [cf. eqs.~\eqref{eq:kpart_meancoh},\eqref{eq:connected}] for $k=4,5$. Indeed, we observe an almost linear relationship between the two quantities over the entire range between indistinguishable fermions and bosons, garnished by small, but systematic, deviations from linearity. Averaging $k$-point correlators over randomly sampled unitaries $U$ thus yields a valid estimate for the corresponding $\mathcal{W}^{(k)}$. This approach is especially promising in reconfigurable linear optical networks \cite{shadbolt_generating_2012,russell_direct_2017,mennea_modular_2018}. Furthermore, Fig. \ref{fig:corvsmeancoh} shows that replacing the random matrix integration by an average over all 
connected correlators $\mathcal{C}^{(k)}_{p_1,\dots,p_k}$ of $k$ out of $d$ output modes for a \textit{single} random unitary leads to a similar linear relation. Note, in this case, that while the resulting slope of $\mathcal{C}^{(k)}_{\bf p}$ vs. $\mathcal{W}^{(k)}$ depends on the specific unitary, deviations from linearity are centered on the predictions from the Haar integration. Indeed, the equivalence of mode average and random matrix prediction is reasonable for large systems: Then, the matrix elements of a (sub-matrix of a) random unitary $U$ are approximately i.i.d. Gaussian, and the mode average realizes a sample mean of the true distribution, which, hence, converges for large samples, by the law of large numbers. This allows to estimate $\mathcal{W}^{(k)}$ in experimental situations where sampling many random Haar unitaries is not possible. 

In \cite{walschaers_statistical_2016,walschaers_many-particle_2016,giordani_experimental_2018,flamini_2020} a statistical analysis of the moments of the distribution of connected two-point correlators (\ref{eq:connected}) was suggested as a certification tool for MBI.
Based on this observation, two-point correlations were also put forward to witness non-classicality \cite{rigovacca_nonclassicality_2016} or indistinguishability \cite{van_der_meer_experimental_2021}. Similar in spirit, the characterization of $N$-photon coherence by the pair-wise overlaps of the particles' internal states in \cite{brod_witnessing_2019,giordani_experimental_2020,giordani_witnesses_2021,galvao_quantum_2020} addresses only two-particle correlations. It is clear from (\ref{eq:kPC},\ref{eq:2pointcor}) that such protocols only yield marginal 2P information contained in $\rho\inext^{(2)}$. However, Fig. \ref{fig:meancoh_bosfer} shows that, for externally separable states, as mostly considered in the literature, higher-order coherence depends monotonically on $\mathcal{W}^{(2)}$.

\paragraph*{External entanglement ---}

For externally entangled states, however, $2$P coherences do not convey unambiguous information on higher-order coherence, as we now demonstrate by example: Take orthogonal external and internal modes $p,q,r$ and $\alpha,\beta,\gamma$, respectively. The entangled 2P state
\begin{equation}\label{eq:2asymstate}
\ket{\psi_2} = \frac{1}{\sqrt{2}}  \big( \hat{a}^\dagger_{p\alpha} \hat{a}^\dagger_{q\beta}
-\hat{a}^\dagger_{p\beta} \hat{a}^\dagger_{q\alpha}
\big)\ket{0} 
\end{equation}
has $\mathcal{W}^{(2)}=0$ for bosons and $\mathcal{W}^{(2)}=2$ for fermions, i.e. the exact opposite of what is obtained for an externally separable state of indistinguishable particles (recall Fig.~\ref{fig:corvsmeancoh}). Such swapping of quantum statistics induced by entanglement has, e.g., been discussed in \cite{michler_interferometric_1996,sansoni_two-particle_2012,matthews_observing_2013}. A further example is given by the entangled 3P state 
\begin{equation}\label{eq:3cyclicstate}
\ket{\psi_3} = \frac{ \hat{a}^\dagger_{p\alpha} \hat{a}^\dagger_{q\beta}\hat{a}^\dagger_{r\gamma}
	+\hat{a}^\dagger_{p\gamma} \hat{a}^\dagger_{q\alpha}\hat{a}^\dagger_{r\beta}
	+\hat{a}^\dagger_{p\beta} \hat{a}^\dagger_{q\gamma}\hat{a}^\dagger_{r\alpha} }{\sqrt{3}}  \ket{0} \,,
\end{equation}
with $\mathcal{W}^{(2)}=1$, but $\mathcal{W}^{(3)}=3$, for both bosons and fermions, which contradicts the strict monotonic dependence between mean coherences of different orders for externally separable states displayed in Fig.~\ref{fig:meancoh_bosfer}. Non-classical correlations as those inscribed into $\ket{\psi_3}$ result in \textit{pure 3P interference}: All 2P coherences $\bra{m,n} \rho\inext^{(2)} \ket{n,m},\, m\neq n$, vanish, such that any two-point correlation, in fact \textit{any} 2P observable as defined in \cite{brunner_signatures_2018,dufour_many-body_2020}, must yield a classical result, while an arbitrary three-point correlator will unveil the coherences of $\ket{\psi_3}$.
States displaying \textit{pure $k$P interference} can be obtained by a suitable generalization $\ket{\psi_3}$. In these states, coherence is exclusively concentrated on the highest order, such that the system
behaves alike classical particles in all measurements of order $k < N$.
This is in contrast to the states introduced in \cite{shchesnovich_collective_2018},
which carry an $N$P phase visible only in highest order ($N$-point)
correlation measurements, but also display lower-order coherence.
Note that states with a cyclic structure similar to \eqref{eq:3cyclicstate} are employed in \cite{karczewski_genuine_2019} to define the notion of \textit{genuine $k$-partite indistinguishability}. Such a phenomenology is realizable only through entanglement and is reminiscent of that of GHZ
states \cite{horodecki2009}.

\paragraph*{Conclusion ---}
The $k$P mean coherence $\mathcal{W}^{(k)}$ of a possibly entangled $N$P state $\rho$ is experimentally directly accessible through the protocol of randomized correlation measurements, see Fig.~\ref{fig:corvsmeancoh}. For externally separable states, $\mathcal{W}^{(2)}$, inferred from two-point correlation measurements, contains already all relevant information about the full state's mean coherence, as shown by the narrow monotonic dependence of $\mathcal{W}^{(k)}$, $k>2$, on $\mathcal{W}^{(2)}$ in Fig.~\ref{fig:meancoh_bosfer}. Any significant deviation of $\mathcal{W}^{(k)}$ from this provides a strong indication of external entanglement. The estimation of $\mathcal{W}^{(\ell)}$, $\ell\leq k$, involves $\binom{d}{k} \sim d^k$ randomized $k$-point correlation measurements. Although the necessary number-resolution is experimentally challenging to implement, our protocol shows a tremendous advantage over estimating the full output counting statistics, which scales exponentially in $N$, and promises diagnostic power to assess the multi-partite entanglement properties of the input state in its external dof.

\begin{acknowledgments}
We are indebted to Andreas Ketterer, Christoph Dittel and Mattia Walschaers for enlightening discussions.
\end{acknowledgments}


\begin{thebibliography}{72}
\bibitem [{Not({\natexlab{a}})}]{NoteX}%
  \BibitemOpen
  \href@noop {} {} \bibinfo {note} {This reduction, first,
  to only external dof, and, second, to a subset of particles, is sensitive to
  non-classical correlations between the external and internal dof, as well as
  between the particles. Such correlations are, however, to be sharply
  distinguished from external mode-entanglement which we here focus
  on.}\BibitemShut {Stop}%
\end{thebibliography}

%

%
%

\newpage

\newcommand{\TITLE}{Many-body coherence and entanglement probed by randomized correlation measurements}
\title{Supplemental Material \\[2mm] \TITLE}


\onecolumngrid
\begin{center}\large\bfseries Supplemental Material \\[2mm] \TITLE \end{center}
\vskip 4mm
\twocolumngrid
\renewcommand\theequation{S\arabic{equation}}
\renewcommand\thefigure{S\arabic{figure}}
\setcounter{figure}{0}
\setcounter{equation}{0}


\subsection{Relation of the mean coherence to quantities defined in the literature}

For Fock states (i.e. eigenstates of all number operators $\hat{N}_{p\alpha}=\hat{a}^\dagger_{p\alpha} \hat{a}_{p\alpha}$), the \textit{degree of indistinguishability}
\begin{align}
\mathcal{I} &=\sum_{m \neq n\in \mathcal{B}\inext}\sum_{\alpha \in \mathcal{B}\inext}  N_{m\alpha}N_{n\alpha}\ \Big/ \! \sum_{m \neq n\in \mathcal{B}\inext} N_{m}N_{n}
\end{align}
was introduced in \cite{brunner_signatures_2018,dufour_many-body_2020} to quantify partial distinguishability in multi-component bosonic systems. In particular, $\mathcal{I}$ was shown to correlate with the time-average of the density variances  $\braket{N_m^2(t)}-\braket{N_m(t)}^2$, which probes the $2P$ reduced state evolving from the initial Fock state. The degree of indistinguishability is related to the 2P mean coherence for arbitrary definite external mode occupations $N_p \in \mathbb{N}$, by
\begin{align}
\mathcal{I}  =N (N-1) (\mathcal{W}^{(2)}-1)\ \Big/ \! \sum_{m \neq n\in \mathcal{B}\inext} N_{m}N_{n}~.
\end{align}

In \cite{brod_witnessing_2019} a notion of  \textit{genuine $N$-photon indistinguishability}, as well as a corresponding witness based on the internal states' overlaps of all pairs of particles, was defined. This witness is further investigated experimentally in \cite{giordani_experimental_2020,giordani_witnesses_2021}. Note that the therein considered class of states is contained in the class of externally separable states with $N_p\in\lbrace 0,1\rbrace$, i.e. mixtures of states where each particle can be associated with a well defined internal state. The witness is derived by noting that for a set of internal states $\ket{\phi_i}, i=1,\dots,N$ with at least two orthogonal states
\begin{equation}
	\sum_{i=1}^N \sum_{j\neq i} |\braket{\phi_i|\phi_j} |^2 \leq (N - 1) (N-2) \,.
\end{equation}
From this inequality we directly obtain an inequality
\begin{equation}
	\mathcal{W}^{(2)} \leq 2 - \frac{2}{N}
\end{equation}
that holds for all states of \cite{brod_witnessing_2019} that do not describe genuine $N$-photon indistinguishable photons. This witness is also experimentally accessible, through our introduced framework of randomized two-point correlation measurements (Eq. (5) and subsequent discussion in the main text).

For an $N$P state $\rho$, the reduced external state $\rho\inext=\rho\inext^{(N)}$ coincides with $1/N!$ times the $J$ matrix introduced in \cite{shchesnovich_tight_2015,shchesnovich_partial_2015} if ideal detectors are assumed. Actually, the author of \cite{shchesnovich_partial_2015} writes ``Note that quantum coherence of photon paths is reflected in the $J$ matrix in a way very similar as in the usual density matrix of a quantum system'' but does not push the connection further. One can measure the bosonic character of the external reduced state $\rho\inext$ by its projection onto the symmetric subspace \cite{shchesnovich_tight_2015,dittel_wave-particle_2019}
\[
p_s=\Tr(\rho\inext P_S) \,,
\]
where $P_S=\frac{1}{N!}\sum_{\pi\in S_N} \pi$ is the symmetrizer and $\pi$ acts on $\bm{m}\in \mathcal{H}\inext^{(N)}$ as $\pi\ket{\bm{m}}=\ket{ m_{\pi^{-1}(1)} ,\dots, m_{\pi^{-1}(N)} }$. This quantity is proportional to the $N$P mean coherence, with
\[
p_s=\mathcal{W}^{(N)}\prod_{m\in\mathcal{B}\inext} N_p!/N! \,.
\]
For particles with individual pure internal states $\ket{\phi_i}$, this is also equal to $1/N!$ times the permanent of the distinguishability matrix  $\mathcal{S}=(\braket{\phi_i|\phi_j})_{i,j}$ introduced in \cite{tichy_sampling_2015}.

\subsection{Sampling of internal states}\label{app:sampling_intstates}

To map out the full transition from indistinguishable fermions to bosons, via the intermediate case of distinguishable particles, in terms of the $k$P mean coherences $\mathcal{W}^{(k)}$ as uniformly as possible, we use the following two-step sampling procedure of pure internal states for each of the particles (the dimension of the internal Hilbert space has to be larger or equal to the number of particles). To sample the neighborhood of indistinguishable particles, we start from a unit vector $\ket{e}\in\mathcal{H}\inint$ and add a perturbation $\ket{f_i}$, with the real and imaginary parts of the components of $\ket{f_i}$ drawn from a normal distribution with zero mean and variance $\epsilon$. By choosing $\epsilon$ sufficiently small, the resulting internal states $\ket{\phi_i}= \ket{e} + \ket{f_i}$, after normalization, are almost parallel. The larger $\epsilon$ gets, the smaller the relative contribution of the constant vector $\ket{e}$ becomes, after renormalization, and we sample the unit sphere in $\mathcal{H}\inint$ almost uniformly. As a second step, we sample the neighborhood of perfectly distinguishable particles by choosing $N$ orthogonal unit vectors $\ket{e_i}\in\mathcal{H}\inint$ (one for each particle) perturbed by vectors $\ket{f_i}$ sampled as before with normally distributed components in $\mathbb{C}$, followed by renormalization. As before, for large $\epsilon$ the contributions from the constant vectors $\ket{e_i}$ in $\ket{\phi_i}= \ket{e_i} + \ket{f_i}$ are negligible, and we approach uniform sampling of the unit sphere in $\mathcal{H}\inint$. For sufficiently small $\epsilon$, we generate states $\ket{\phi_i}$ in the vicinity of perfect distinguishability. This procedure is followed for fermionic and bosonic particles. In both cases the limits of distinguishable particles coincide, with $\mathcal{W}^{(k)} =1$ for all $k\leq N$.

\end{document}